# AGENT-BASED MODEL FOR TUMOR-ANALYSIS USING PYTHON+MESA

**Ghazal Tashakor [(a)], Remo Suppi [(a)]**

(a) Universitat Autònoma de Barcelona, Department Computer Architecture & Operating Systems, School of Engineering, Campus Bellaterra,
Cerdanyola del Vallès, Barcelona, 08193, Spain

[(a)] Ghazal.Tashakor@caos.uab.cat  [(a)] Remo.Suppi@uab.cat

**ABSTRACT**
The potential power provided and possibilities presented by computation graphs has steered most of the available modeling techniques to re-implementing, utilization and including the complex nature of System Biology (SB). To model the dynamics of cellular population, we need to study a plethora of scenarios ranging from cell differentiation to tumor growth and etcetera.
Test and verification of a model in research means running the model multiple times with different or in some cases identical parameters, to see how the model interacts and if some of the outputs would change regarding different parameters.
In this paper, we will describe the development and implementation of a new agent-based model using Python. The model can be executed using a development environment (based on Mesa, and extremely simplified for convenience) with different parameters. The result is collecting large sets of data, which will allow an in-depth analysis in the microenvironment of the tumor by the means of network analysis.

## 1. INTRODUCTION

At the cellular level, Systems Biology (SB) is mainly a complex natural phenomenon that today has become important for research in certain areas such as drug development and biotechnological productions and applications.
Mathematical models of cellular level are built in bio repeated cycles at a portion of time to arrive at a decision. The primary idea behind the bio iterative cycles is to develop systems by allowing software developers to redesign and translate the cellular metabolism. The objective is bringing the desired decision or result closer to discovery in each iteration; these experiments on multi-scale iterative cycle of computational modeling -besides experimental validation and data analysis- would produce incremental samples for the high-throughput technologies.
Also the behavior of the cell emerges at the network-level and requires much integrative analysis. Moreover, due to the size and complexity of intercellular biological networks, computational model should be a considerate part of the production or application.
In the following whole simulation challenges, SB would not be needless of developing integrated frameworks for analysis and data management. Alongside the intercellular level, many researches in SB also address the issue of cellular population.
In evolutionary versions of these scenarios, modeling the interactions of these type of models across large multi-scale would need agent-based modeling. Each agent has a Boolean network for its own expression, much like a gene. So a proper production or application for SB must support not only a compatible simulation method but also suitable methods for model parameter estimations which represent the experimental data (D.2011; An2008).
Therefore, ABMs start with rules and mechanisms to reconstruct through the mathematical or computational form and observe the pattern of data. Processing the heterogeneous behavior of individual agent within a dynamic population of agents -which cannot be controlled by an overall controller- needs a higher-level system parallelism which ABMs supports.
Biological systems include random behaviors and ABMs accommodate this via generation of population agents in the agent's rules.
ABMs have a level of abstraction to create new cellular states or environmental variables without changing core aspects of the simulation. To aggregate the paradoxical nature of emergent behavior which could be observed from any agent in contrast to a conceptual rules of the model, ABMs reproduce emergent behavior. Emergent behavior has a range of stochasticity similar to real world Systems Biology (Bonabeau2002).
Finding software platforms for scientific agent-based models require comparing certain software design parameters such as emulating parallelism, and developing schedulers for multiple iterations, which manage ABM run.
Many references reviewed and compared different agent-based modeling toolkits. However, from the perspective of biotechnological application and biotechnologists, most of them share a key weakness. It is using complex languages which are not Python. Perhaps re-implementing ABMs in Python would be a wiser technical strategy since it is becoming the language of scientific computing, facilitating the web servers for direct visualization of every model step, debugging and developing an intuition of the dynamics that emerge from the model, also allowing users to create agent-based models using built-in core components such as agent schedulers and spatial grids (Villadsen and Jensen 2013).

## 2. BACKGROUND

Spontaneous tumor, which progresses from the initial lesion to highly metastatic forms are generally profiled by molecular parameters such as prognosis response, morphology and pathohistological characteristics. Tumors can induce angiogenesis and lymph-angiogenesis, which plays an important role in promoting cancer spread. Previous studies have shown that the cancer stem cell (CSC) theory could become a hypothesis for tumor development and progression.

These CSCs have the capability of both self-renewal and differentiation into diverse cancer cells, so one small subset of cancer cells has characteristics of stem cells as their parents. Hereditary characteristics play a certain role in malignant proliferation, invasion, metastasis, and tumor recurrence.

In recent researches the possible relationship between cancer stem cells, angiogenesis, lymph-angiogenesis, and tumor metastasis is becoming a challenge. Due to many evidences and reviews such as (Li 2014; Weis and Cheresh 2011; Carmeliet, P., & Jain, R. K. 2000), metastasis is defined as the spread of cancer cells from the site of an original malignant primary tumor to one or more other places in the body. More than 90% of cancer sufferings and death is associated with metastatic spread. In 1971, Folkman proposed that tumor growth and metastasis are angiogenesis-dependent, and hence, blocking angiogenesis could be a strategy to intercept tumor growth.

His hypothesis later confirmed by genetic approaches. Angiogenesis occurs by migration and proliferation of endothelial cells from original blood vessels (Weis and Cheresh 2011).

Accordingly, translational cancer research has contributed to the understanding of the molecular and cellular mechanisms occurring in the tumor and in its microenvironment which causes metastasis and this could present a model relatively similar to physiology of human or at least has the capability of going through genetic manipulations that bring them closer to humans. Hence tumor modeling with a high spatiotemporal resolution combined with parametric opportunities has been rapidly applied in technology (Granger and Senchenkova 2010).

### 2.1. Agent-based Models in Systems Biology

A primary tumor model addressing the avascular growth state depends on differential equations. They are classified as "lumped models" to predict the temporal evolution of overall tumor size. Since lumped models just provide a quantitative prediction of tumor size over time with only a few parameters and very low computational results, they would not be enough for an explicit investigation of many other events such as spatiotemporal dynamics of oxygen and nutrients or cell to cell interactions; also, stromal cells which play a major role in cancer growth and progression in the interaction with tumor cells. The result is disregarding the mutations in the tumor microenvironment and metastasis.

These shortcomings lead us to In Silico models of tumor microenvironment. In Silico (Edelman, Eddy & Price 2010) refers to computational models of biology and it has many applications. It is an expression performed on a computer simulation. In Silico models are divided to three main categories (Thorne, Bailey, & Peirce 2007; Soleimani et al., 2018):

1. Continuum based models which solve the spatiotemporal evolution problems of density and concentration of cellular population in the tumor microenvironment.
2. Discrete or agent based on a set of rules which change the cells' states and manage the cells' interactions within the tumor microenvironment.
3. Adaptive Hybrid models which integrate the above solutions.

Mathematical and computational models of the tumor should cover different aspects of tumor interactions in its microenvironment. Most of the avascular tumor models address the tumor growth dynamics not taking into account the angiogenesis setting; but computational models have started research on tumor-induced angiogenesis since the beginning of cancer research.

Therefore, the angiogenesis models have a very strong background in experimental observations. The most established researches have explicitly focused on the stromal cells to achieve a close computational model of angiogenesis.

For example, according to Monte Carlo simulations and energy minimization, cellular models are expanded and elaborated on cellular automata to allocate more than one lattice site to each cell and describe cell to cell and tumor stroma interactions. Nevertheless, building a bulky model over a range of matrix densities -which covers numerous factors for a large domain size and 3D simulations- is restricted by computational and application costs.

Based on a paper written by Soleimani et al. (2018), a minimal coupling of a vascular tumor dynamics to tumor angiogenic factors through agent-based modeling has progressed the experimental studies in the recent years. They have mentioned in both reviews that it is challenging to mathematically simulate all the process of a complex system such as tumor growth, metastasis and tumor response treatments, because mathematical modeling is still a simplification of the system biology and the results requires validation.

Also, building a predictive experimental plus theoretical application without clinical data requires parametrizing and validating again. The best improvement which is expected would be predictive modeling in both preclinical and clinical states (Thorne, Bailey, & Peirce 2007).

## 2.2. Complex Networks in Biological Models

Another interesting approach is the network models. Networks follow patterns and rules and have a specific topology, which allows scientists to go through with a deeper investigation towards biology information extraction.

Within the fields of biology and medicine, Protein-protein interaction (PPI) networks, biochemical networks, transcriptional regulation networks, signal transduction or metabolic networks are the highlighted network categories in systems biology which could detect early diagnosis.

All these networks need compatible data to be produced experimentally or retrieved from various databases for each type of network; but besides analyzing data structures for computational analysis, several topological models have been built to describe the global structure of a network (Girvan, Newman 2002; 2004).

## 3. MODELING AND SIMULATION

Our approach to the generation of an ABM model considered different temporal and spatial scales, focusing first on mitosis as the main axis of tumor growth. This model and its analysis developed in Netlogo (section 3.1). Then large number of environment limitations were detected with a high number of agents and difficulties to analyze the microenvironment of emergent growth. The second approximation is presented in this paper (section 3.2) used dynamic networks based on a large number of interactive agents. This model makes it possible to help researchers in carrying out more detailed research on intercellular network interactions and metastasis in a multiple scale model (Grimm2005).

## 3.1. Netlogo Model and Experimental Results

Our ABM NetLogo model is designed as a self-organized model that illustrates the growth of a tumor and how it resists chemical treatment.

This model in NetLogo (based on Wilensky's tumor model (Wilensky 1999)) permits us to change the parameters that affect tumor progression, immune system response, and vascularization. Outputs included the number of living tumor cells and the strength of the immune system which control cells. In this model the tumor has two kinds of cells: stem cells and transitory cells. In the model, tumor cells are allowed to breed, move, or die.

The simulation presented cells' control with different and constant immune responses through killing transitory cells, moving stem cells and original cells.

Figure 1 shows the steady state of a tumor metastasis visualization with 6 stem cells and the grow-factor =1.75, replication-factor=high, and apoptosis=low.

As it could be seen, the growth of metastasis is more aggressive and through reducing apoptosis, there is a greater number of cells that do not die, amounting to near 200,000 cells (agents) (Tashakor, Luque & Suppi, 2017).

The main problem with this implementation was the limitations of the execution environment (Java memory limitations) and loss of performance with a high number of metastasis cells. In addition, this model did not allow capturing in detail the interactions between the different parameters at the microenvironment level.

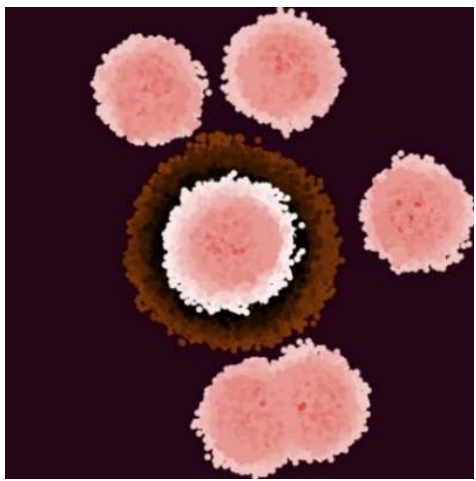

Figure 1: stem cells evolution and metastasis visualization with grow-factor=1.75, apoptosis=low and replication-factor=high (near 200,000 cells in steady state).

## 3.2. Python Model and Experimental Results

To solve the aforementioned problems, we started a new development using Python +Mesa. Mesa is the agent-based Python project, which has started recently and it has rapidly found its place among the researchers. Mesa is modular and its modeling, analysis and visualization components are integrated and simple to use.

This feature convinced us to re-implement our tumor model with Mesa network structure. Furthermore, it supports multi-agent and multi-scale simulations, which is suitable for creating dynamic agent-based models. This paper documents our work with the Python +Mesa to design an agent model based on the tumor model scenario.

At the beginning, Mesa was just a library for agent-based modeling in Python but now it is a new open-source, Apache 2.0 licensed package meant to fill the gap in the modeling complex of dynamical systems.

In Mesa, the Model class to define the space where the agents evolve handles the environment of an agent-based model. The environment defines a scheduler which manages the agents at every step.

For studying the behavior of model under different conditions, we needed to collect the relevant data of the model while it was running; Data Collector class was defined for this task. The adapting modules of Mesa allow us to make changes in the existing ABMs for the purpose of conforming with the requirements of our future framework.

Also, monitoring the data management issues when processing actions happen in parallel seems facile in Mesa, since each module has a Python part which runs on the server and turns a model state into JSON data (Masad & Kazil, 2015). The advantage of Mesa is its browser for visualization which allows the user to see the model while running in the browser.

Tumor progression is a complex multistage process and the tumor cells have to acquire several distinct properties either sequentially or in parallel.

The first problem using NetLogo framework for simulation was using behavioral space which supports multithreading. In this way, performance is limited to the number of cores at the local infrastructure.

To solve the local processing problem, we utilized the parametric simulations using a HPC cluster in order to reduce the necessary time to explore a determinate model data space.

Since we needed to distribute our model through the distributed architecture to explore the model states in the microenvironment scale, it was revealed that NetLogo also has scalability issues in designing graph networks.

Although there are available extensions for scaling up the model in the form of graph, it still depends on the size of the tumor and it could be a very slow process, taking days to finish experiment considering the checkpoints and performance bottlenecks while using a HPC cluster.

We had to translate and move our model to a new framework such as the style and structure of Mesa to facilitate our distributed executions.

The newest changes in the ground of the old scenario is increasing the population of agents, obtaining graph-based representation of biological network and consideration of multi-state and multi-scale components (An2008).

The initial design aspect of the multi-scale architecture of out tumor growth model in Python comes from the acute inflammation based on the key factors such as angiogenesis. Tumor angiogenesis is critical for tumor growth and maintenance (Kaur et al., 2005).

Our new model strategy begins with an initial identification of a minor population of cells with the characteristics of "tumor-initiating" cancer stem cells and these cancer stem cells in the assumed tumor reside in close proximity to the blood vessels. Therefore, we chose an angiogenetic switch for our model which have to balance this dynamic. We had angiogenetic and anti-angiogenetic factors.

In the case of anti-angiogenetic, we will have the probability of quiescent tumor. In this agent-based model, tumor cells are affected, inflamed and turn quiescent.

Table 1 shows the range of parameters we implemented to control the factors.

Based on these factors, we can achieve a simulation evolution of metastasis and measure the tumor volume ratio along with the dynamics of the tumor under the influence of the factors.

Table 1: different control factors of tumor growth

| Factors | | | |
|---|---|---|---|
| | Low | Medium | High |
| Angiogenesis | 0.0 | 0.4 | 1 |
| Recovery | 0.1 | 0.3 | 1 |
| Quiescent | 0.1 | 0.5 | 1 |

To select a network topology for tumor model, we selected a random graph. Barabasi-Albert model is a scale-free network and it is one of the most basic models since it describes most of the biological networks - especially in evolutionary models; but since there is a need to manage the cell interactions and stromal cells behavior within the tumor microenvironment and due to the time-dependency of the connections, we developed our Python agent-based model on Erdös-Rényi topology. The visualization of the model is a network of nodes (using the Mesa architecture) that shows the distribution of agents and their links. Considering the architecture of Mesa, the creation of agents occurs based on the assignment of nodes to a graph. A scheduler (time module) activates agents and stores their locations and updates the network.

The total operation time is directly related to the number of steps necessary to deploy all the agents. Since each agent changes in three states, the process goes on until the tumor agent's volume appears as metastasis.

Thus, for P as a Probability and for edge creation in n number of nodes, if $(Ps > log(n)/n))$ almost all vertices are connected and this function returns a directed graph as we described in Algorithm 1.

Erdös-Rényi model takes a number of vertices $N$ and connecting nodes by selecting edges from the $(N(N-1)/2)$ possible edges randomly. The pseudo-code generating a random network is described in Algorithm 1:

```
Algorithm 1
Input: (n,p, p* )
Output: (True, False)
Begin Algorithm
  For n in enumerate(nodes):
    For p in enumerate(Ps):
      If  p > p* then
          p_connected(n, p) = 1
      Else
          return 0
      Endif
    Endfor
  Endfor
End Algorithm
```

The interactive visualization in Mesa helps us to identify insights and generate value from connected data. Considering the data analysis for life sciences such as tumor which is almost about connections and dependencies, the large amount of data makes it difficult for researchers to identify insights.

Graph visualization makes large amounts of data more accessible and easier to read, as it has been illustrated in Figure 2.

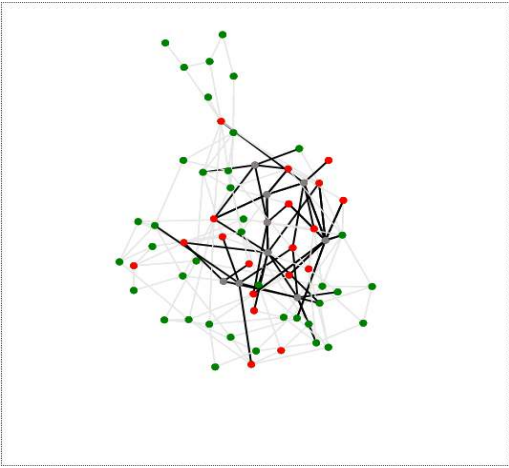

Figure 2: Graph visualization for three states (normal, dead and metastatic) shown in three colors

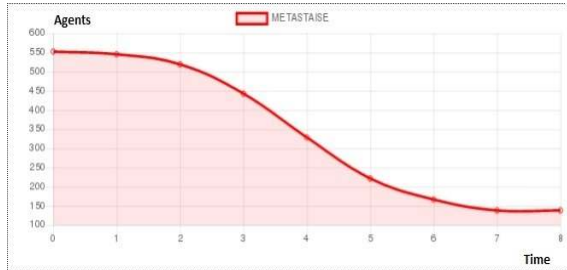

Figure 3: Tumor stabilization challenges for cancer stem cells (CSC) with control factors more than medium

Figure 3 shows the reduction of cancer stem cells in 550 initial stem cells which are affected by control factors in the time it took, which was about 8 minutes for this simulation.

Tumor growth curves are used for derivation of the Tumor Control Index (TCI), such as tumor progression, rejection and stabilization. We tried to combine both rejection and stabilization challenges to show the potential of Python agent-based modeling with different control factors.

According to (Peskin 2009), we need a statistical method that leads to volume ratio measurements of tumor, as it could be seen in figure 3. To calculate tumor volume using (Monga.s 2000) we needed tumor width (W) and tumor length (L) which is presented in formula 1:

$$Tumor\ Volume = W^2 \times \frac{L}{2} \qquad (1)$$

This calculation which works pretty well for clinical issues, was made based upon the assumption that solid tumors are more or less spherical like the version we had before in Netlogo Wilensky's tumor model, but not proper for the metastatic diseases, which due to the phase transition and spreading dynamics will be defined by graph presentations in the future.

According to Rai (2017), the model which have the same size and number of connections as of a given network could maintain the degree sequence of the given network. By generating a random network with a given average degree (K) and initial size of tumor (N), we could construct degree sequence (*m*) and it is presented in formula 2:

$$m = \frac{1}{2}\sum_{i=1}^{N} Ki$$
$$N = \binom{n}{2},\ K=\{3,4,5,6,7,8\} \qquad (2)$$

For volume calculation, we start by assuming that every cell inside the tumor has three states (normal, dead and metastatic) which corresponds to tumor by edges. With the given K degree from three to eight and N initial nodes, we constructed an ER (Erdös-Rényi) network, whose degree sequence of (*m*) could lead us to the tumor volume ratio.

The pseudo-code using the dynamic of degree sequence under the influence of above factors producing tumor volume in graph network is described in Algorithm 2:

| Algorithm 2 |
|---|
| **Input**: (N,K=3) |
| **Output**: (Tumor Volume) |
| Begin Algorithm |
|    While  p_connected= 1 |
|         For N in nodes(G) |
|             If P $\binom{N}{K}$ < Factor then |
|                   Return m/n |
|             Endif |
|         Endfor |
| End Algorithm |

In Figure 4, we executed the model with different angiogenesis control factor in (0.1 to 0.9) for 60, 360, 650, 100 and 1200 cancer stem cells (CSCs) to produce the tumor volume analysis based on the density distribution of the graph.

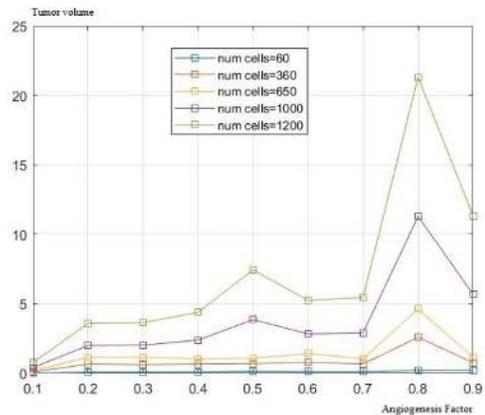

Figure 3: Tumor Angiogenesis volume ratio for 60 to 1200 cancer stem cell (CSC)

## 4. CONCLUSION

In summary, agent-based modeling in Python with Mesa project represents a valuable and time-saving simulation technique, which allowed us to re-implement our tumor model across a network based space.

Mesa enable a modeler to quickly write down the code and quickly explore the results. Mesa also collect the results easily and help us in data integration which is very important in the case of phase transition and pattern-oriented modeling.

We migrated successfully from Netlogo to Python-Mesa, which will allow us to reintegrate our system quicker and reiterate simulations taken on previous system with more data in considerably less time.

In the future, we will extend the model to run on cloud infrastructure in parallel. This could be an impressive achievement for fast analysis purposes in clinics, both on the predictive diagnostic and therapeutic side.

As advantages of this contribution, we have a scalable model that together with Python can be distributed over an HPC architecture eliminating the limitations of other environments (e.g. Netlogo due to memory limitations of the JVM).

When Python + Mesa can be deployed over an HPC architecture, there will be a notable increase in scalability and performance since, in this type of environment, the simulation will not be limited by memory.

One of the most important disadvantages is the model visualization and animation for oncologists. Netlogo has a cells visualization and animation that can be useful to medical environments. Mesa only has the visualization of the network of agents and their interconnections, so graphic functions must be developed to see the evolution of the cells and the degree of metastasis/apoptosis and other relevant parameters for the medical analyst.

Another important aspect that must be addressed will be the functional validation of the whole model since at this moment only a verification and partial validation of the behavior of the affected cells has been carried out.

**ACKNOWLEDGMENTS**

The MINECO Spain under contract TIN2014-53172-P and TIN2017-84875-P have supported this research.

**AUTHORS BIOGRAPHY**

**GHAZAL TASHAKOR** is Research Fellow and PhD Candidate for trainee research staff position (PIF) at the Computer Architecture & Operating Systems Department (CAOS) of the University Autònoma de Barcelona. Her research interests include high performance simulation, agent-based modelling and their applications, service systems, integration, resource consumption, and execution time. Her e-mail address is: ghazal.tashakor@caos.uab.cat

**REMO SUPPI** received his diploma in Electronic Engineering from the Universidad Nacional de La Plata (Argentina), and the PhD degree in Computer Science from the Universitat Autonoma de Barcelona (UAB) in 1996. At UAB, he spent more than 20 years researching on topics including computer simulation, distributed systems, high performance and distributed simulation applied to ABM or individual-oriented models. He has published several scientific papers on the topics above and has been the associate professor at UAB since 1997, also a member of the High Performance Computing for Efficient Applications and Simulation Research Group (HPC4EAS) at UAB. His email address is: remo.suppi@uab.cat.